\documentclass{elsart}
\journal{New Astronomy}

\usepackage{natbib}

\usepackage{graphicx}

\usepackage{amssymb}

\begin{document}

\begin{frontmatter}


\title{The Atacama Cosmology Telescope Project: A Progress Report}


\author{Arthur Kosowsky}
\address{Department of Physics and Astronomy, University of Pittsburgh, Pittsburgh, PA 15260, USA}
\ead{kosowsky@pitt.edu}
\collab{for the ACT Collaboration}

\begin{abstract}
The Atacama Cosmology Telescope is a project to map the microwave background radiation at arcminute angular resolution and
high sensitivity in three frequency bands over substantial
sky areas. Cosmological signals driving such an experiment
are reviewed, and current progress in hardware construction is
summarized. Complementary astronomical observations in other
wavebands are also discussed. 
\end{abstract}

\end{frontmatter}

\section{Science Motivation and Sky Signals}

The full-sky microwave maps from the WMAP satellite have provided a definitive measurement of the microwave background temperature on angular scales down to around 15 arcminutes, in five
frequency bands from 23 to 90 GHz \cite{ben03}. These measurements determine the primordial
power spectrum of fluctuations in the baryon-photon plasma of the early universe, 
resulting in tight constraints
on the basic cosmological model of the universe \cite{wmap_params}  and bringing to fruition a decade of anticipation
\cite{jun96}. While additional measurements will provide refined constraints
on some cosmological parameters, they are unlikely to alter qualitatively our fundamental picture
of the universe which has emerged.

Significant experimental attention has now turned to microwave background temperature
measurements at smaller angular scales. At scales smaller than 10 arcminutes, the power
in primary fluctuations produced at the surface of last scattering begins to drop precipitously,
due to diffusion damping on length scales smaller than the thickness of the last
scattering surface. But secondary fluctuations in microwave temperature at arcminute angular
scales are induced by interaction of the microwaves with large-scale structure at redshifts
between zero and a few. In constrast with large-angle temperature measurements, which
image fluctuations in the early universe, high-resolution temperature measurements mainly probe the subsequent development of structure in the universe. The Atacama Cosmology Telescope (ACT)
project \cite{kos03,fow05,act_site} is one effort to map the microwave background at angular resolutions better than two
arcminutes, at sensitivities sufficient to detect all major effects contributing to the temperature fluctuations at these scales. The rest of this introductory section will review the specific signals of interest; then
an overview of the ACT experiment and its current status will be given, along with plans for
complementary astronomical observations in other wave bands. 

\subsection{Power Spectrum}

The power spectrum of primordial perturbations, usually characterized by the power law
index $n$, has been measured by WMAP over a range of multipole moments from
$l=2$ to $l=700$ \cite{wmap_params}.  Further constraints on $n$ come from combining WMAP data
with other measures of power at smaller scales (e.g., \cite{sel05}), but these heterogeneous approaches
are susceptible to systematic errors; the microwave background provides the cleanest
way to measure the power spectrum. ACT will extend the range of precision multipole
measurements out to the point where the primary fluctuations cease being the
dominant contribution to the temperature fluctuations, around $l=3500$.  In principle,
ACT will be able to measure $n$ over nearly an additional decade in scale, 
significantly increasing the precision with which $n$ can be constrained. Measurements
over a larger range of angular scales also increase sensitivity to any departure
from a perfect power law, often characterized by the ``running'' of the spectral
index $dn/d\ln k$ \cite{kt95}.  Determination of $n$ and its running are the most powerful
current sources of information  about inflation or other fundamental processes
producing fluctuations in the very early universe. In practice,
the precision in $n$ will be determined by the accuracy of the relative calibration of
ACT maps at 145 GHz with WMAP maps at 90 GHz and much lower resolution. We
are currently studying the issue of relative calibration using sky and instrument
simulations outlined below.

\subsection{Gravitational Lensing}

Mass density variations along the path of microwave photon propagation result in
the photon experiencing a varying gravitational potential, and thus deflections
in the direction of the potential gradient. This gravitational lensing effect is always
small in cosmological contexts; the propagation direction of an average microwave photon will be
deflected by an angle of tens of arcseconds over its journey of 13 billion light years. 
These deflections are also coherent over larger scales up to a degree, so that
detecting them requires more work than simply having angular resolution better than
the detection angle. The original temperature fluctuations are a gaussian
random field to a good approximation; gravitational lensing will induce a particular
additional four-point correlation function into the temperature map \cite{ber97}. Detecting these
characteristic correlations requires comparing temperature fluctuations in the maps
at small scales of a few arcminutes with larger scales around a degree \cite{oka02,kes03,hir03}. 
The overall
amplitude of the temperature signal is on the order of the mean temperature difference
on the deflection scale, which is on the order of five $\mu$K. 

In addition to lensing from large-scale structure distributed evenly across the sky,
galaxy clusters also give localized gravitational lensing signals. If the cluster
is in front of a background microwave temperature gradient, the lensing produces
a dipole-like pattern aligned with the temperature gradient \cite{sel00}. The separation of the
hot and cold peaks is on the order of one arcminute and the peak-to-peak amplitude
can be of order 10 $\mu$K for a large cluster in front of a substantial gradient \cite{hol04,dod04,val04}. Such signals are in principle observable with ACT's sensitivity and resolution, but
other signals associated with clusters will make reliable lensing mass estimates
difficult.

\subsection{Thermal Sunyaev-Zeldovich Effect}
The thermal Sunyaev-Zeldovich (SZ) Effect describes the energy boost received by
low-energy microwave background photons Compton scattering from high-energy electrons
in hot, ionized gases \cite{zel69}. The SZ effect has so far been detected towards a number of galaxy clusters through pointed observations in the direction of known clusters (see \cite{car02} for
a review);
ACT and similar experiments aim to detect thousands of clusters via blind surveys of
large sky areas. The effective temperature shift of the microwave background blackbody
spectrum has the characteristic form of a Compton y-distortion;
below the ``null" at a frequency of around 217 GHz the radiation flux is decreased, while above
217 GHz the flux is increased, and the amplitude is proportional to the pressure of
the ionized gas integrated along the path of photon propagation. The largest galaxy
clusters have SZ effective temperature distortions as large as 100 $\mu$K, making them the
largest amplitude signals expected at ACT frequencies. 

Unlike clusters selected via X-ray or optical observations, the cluster SZ signal is
essentially independent of cluster redshift for a cluster of a given mass and gas
distribution. The SZ signal, which depends linearly on the electron density, is less sensitive to internal gas structure than the X-ray emission, which is quadratic in the electron density. Selecting
clusters via their SZ distortion is also physically much simpler than selecting clusters optically
via the light from their resident galaxies, since the SZ signal is directly related to the majority
of the baryonic cluster mass, while the optical light comes from galaxies which represent only
a small fraction of the baryonic mass, and whose light emission is due to complex, highly
nonlinear processes which are not entirely understood. These considerations make
thermal SZ distortions an ideal method for selecting galaxy cluster samples of cosmological 
interest. The overall number density of clusters in the universe at a given redshift and
mass limit is a sensitive function of cosmological expansion history \cite{eke96,bar96}, 
making this a potentially
powerful probe of dark energy \cite{wel02}; combining an SZ catalog with optical cluster redshifts provides
a probe of cluster growth through time. The difficulty in this approach is that cluster number
counts also depend sensitively on systematic errors, particularly in understanding the cluster
mass selection function \cite{maj03,fra05}. 
Over the coming few years, significant numbers of clusters with
SZ, X-ray, and optical observations, including weak-lensing mass estimates, will reveal
whether clusters are regular enough to understand systematics in their selection functions
and exploit them as probes of cosmology.

Outside of clusters, small diffuse SZ signals may also come from 
from hot gas in numerous individual galaxies heated by star formation or active galactic nuclei \cite{pla02} or
from the first generation of stars \cite{oh03}.
These signals will be at amplitudes of a few $\mu$K, and the extent to which they can be exploited
for useful scientific information, as opposed to serving only as a noise source, is not
yet clear. 

\subsection{Kinematic Sunyaev-Zeldovich Effect}
The kinematic Sunyaev-Zeldovich Effect also arises from photon-electron scattering; it
is essentially a Doppler distortion coming from bulk velocity of the electrons along the
line of sight, and is proportional to the cluster ionized gas mass times its radial peculiar velocity 
\cite{sun80}. 
For large galaxy clusters with characteristic radial velocities of a few hundred
km/s, the kSZ blackbody distortion will be between a few and 10 $\mu$K. Simulated clusters show that the internal
gas motions in a cluster give an intrinsic scatter of around 50 km/s to cluster velocities
inferred from kSZ observations \cite{nag03}. In principle, galaxy cluster radial peculiar velocities
can be measured directly from kSZ observations, if cluster velocities and gas masses can be disentangled using thermal SZ or X-ray observations \cite{seh05}. The cluster velocity field then provides
a direct and clean probe of the cosmic velocity field and an independent test of the hierarchical structure formation picture. Such velocity catalogs would be a significant improvement over
current velocity catalogs based on galaxy velocities, which are inferred from their redshifts
combined with a distance estimate. 

\subsection{Other Sky Signals}
These sky signals -- thermal and kinematic SZ effect, gravitational lensing, and
primary microwave background blackbody fluctuations, can be separated through
their frequency and spatial dependences. Detecting them will be complicated, however, 
by other astrophysical ``noise" sources (which, however, are interesting in their own right!).
Galactic dust can contribute substantial signal at these wavelengths; to minimize 
this signal, the cosmologically useful portions of the sky are likely to be well away
from the galactic plane. Infrared point sources, which are thought to be high-redshift
galaxies with significant dust emission from massive star formation, will likely give a confusion noise limit in the range of a few $\mu$K at ACT frequencies \cite{kno04,whi04}; removing the brightest infrared point sources through complementary observations at higher redshift will certainly be helpful. Likewise, bright radio point sources, generally emission from active galactic nuclei, can contribute, though these will probably have signficantly less overall signal than the infrared sources, judging from naive extrapolations of microwave and SCUBA observations \cite{kno04}, unless the radio sources are strongly correlated with clusters . 

\section{Telescope}

Observing the signals outlined in the preceding section requires mapping the microwave sky
in several frequency bands with an angular resolution of around an arcminute and noise of
a few $\mu$K per resolution element. These science requirements drive a number of design decisions evident in the ACT telescope. To observe the frequency dependence of the thermal SZ effect
requires frequencies in the neighborhood of 200 GHz; ACT has three frequency bands with
central frequencies (bandwidths) of
147 (23) GHz, 215 (23) GHz, and 279 (32) GHz. These bands take advantage
of atmospheric windows, significantly reducing atmospheric noise. At
these frequencies, the resolution requirements dictate a primary mirror of at least
6 meter diameter; ACT's 6-meter primary 
is constructed from aluminum panels. The need
for control of systematic errors to detect temperature anisotropies which are tiny
fractions of the total signal lead to an off-axis Gregorian optical design with a clear aperture,
a large 2-meter secondary mirror, and an optical path which has no moving
elements, including ground screens which reflect into sidelobes. The focal plane is a square with side 22' on the sky, with a plate scale of approximately 0.7' / mm.

The telescope is designed to
make observations at a fixed elevation while scanning in azimuth, so that the air mass the
microwave photons propagate through is as nearly constant in time as possible: an elevation
change of one degree changes the atmospheric signal by tens of mK, thousands of times
larger than the few $\mu$K sky signals we aim to detect. Scanning in azimuth allows repeated
imaging of the same patch of sky on time scales short enough to overcome detector $1/f$ noise; 
the telescope is designed to complete a scan
with an amplitude of a few degrees in azimuth in 5 seconds. To obtain uniform sky coverage,
the scan should be as close to a constant angular velocity as possible; the acceleration at the
scan turnaround points compared to the optical tolerance determines the necessary structural stiffness
of the telescope structure and the specifications of the hydraulic azimuthal drive motor. The telescope
also has an elevation drive for checking calibration sources or to accommodate observations
at different central elevations. The telescope is being built by AMEC Dynamic Systems in
Port Coquitlam, British Columbia.

\begin{figure}[!t]
\includegraphics[width=13.8cm]{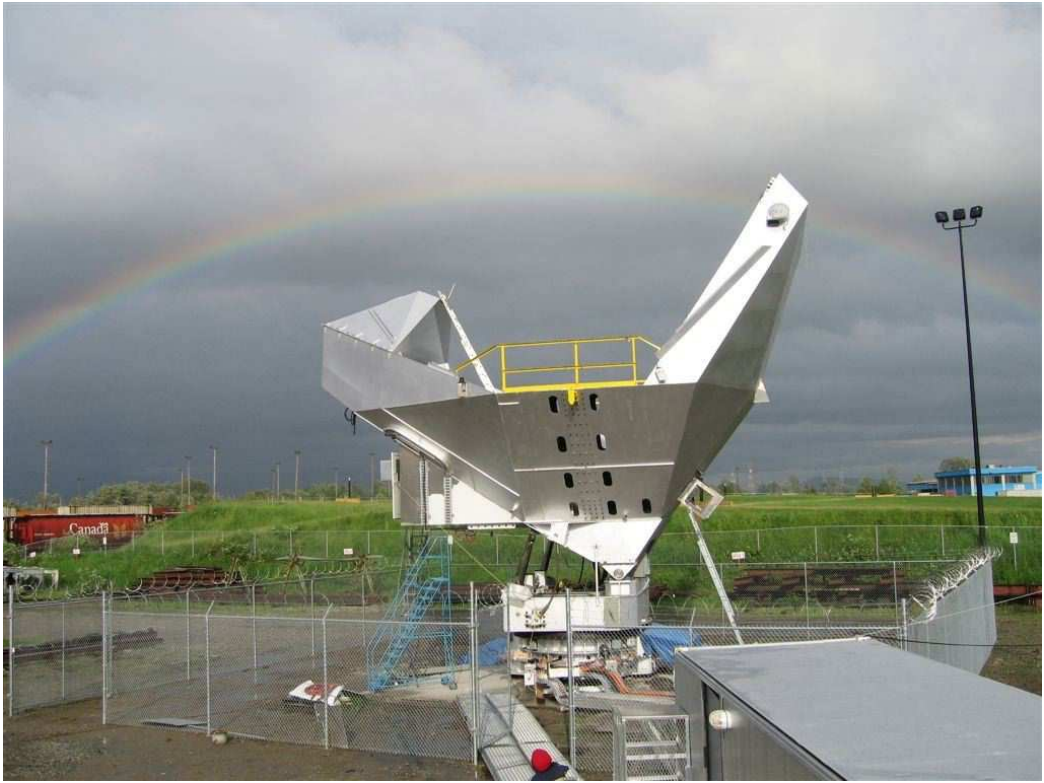}
\caption{The ACT telescope, during construction in Port Coquitlam, British Columbia. 
The 6-meter primary
is contained in the right part of the structure; the 2-meter secondary is at left, and the
detectors are contained in a large cryogenic dewar behind the housing in the center.
The large azimuth bearing on which the entire structure rotates is visible at the bottom. Additional
ground screens attach to the outside of the structure shown here. The top of the primary
housing is 13.5 meters from the ground. It is clear that Port Coquitlam is not an
ideal observing site!}
\label{fig:telescope}
\end{figure}

\begin{figure}[!t]
\includegraphics[width=13.8cm]{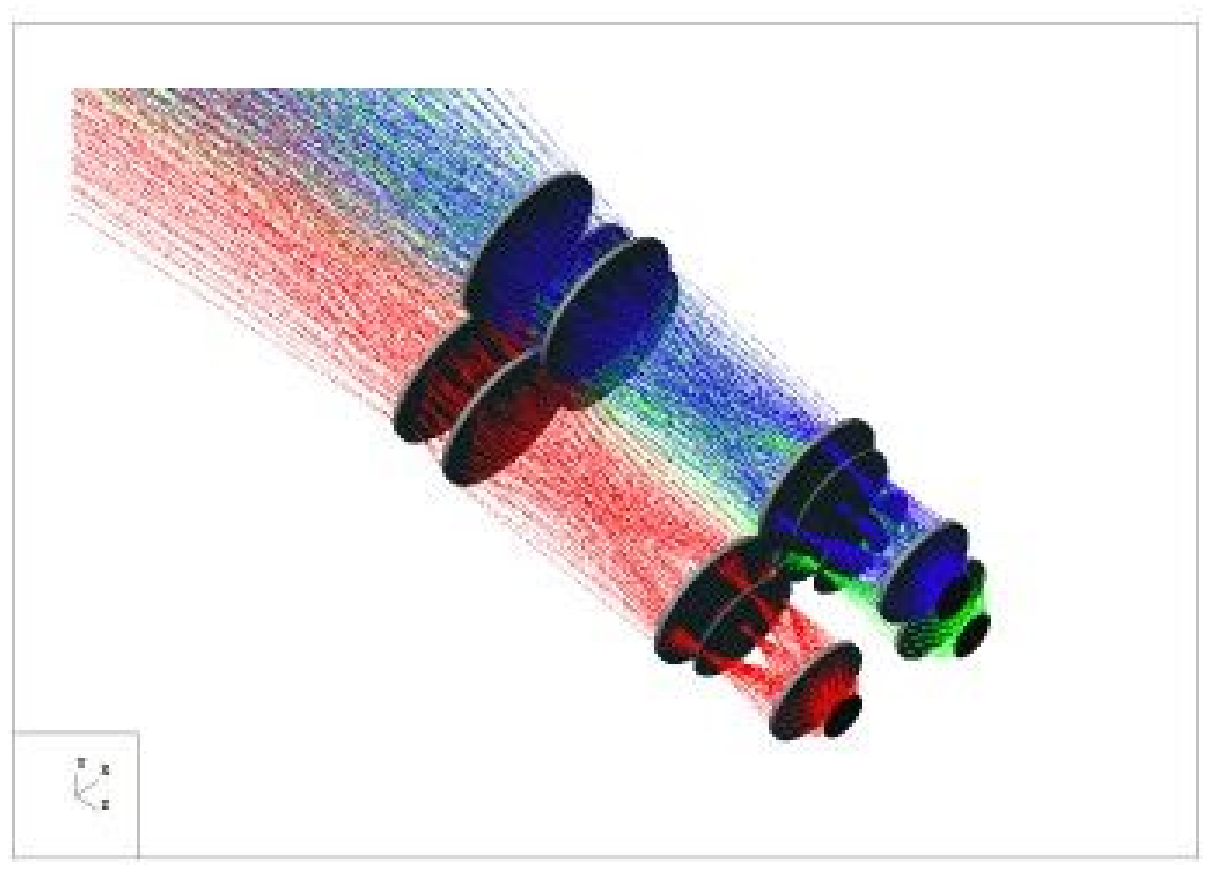}
\caption{The ACT optical design, showing the optical elements in the three optics
tubes. Radiation from the secondary mirror is
coming from the top left; three independent sets of cold filters and lenses
image the three frequency bands onto the focal plane. Mean Strehl ratios 
for point sources are greater than 0.985 in all three frequencies, and
95\% of the focal plane has Strehl ratios higher than 0.96 in all three frequencies.}
\label{fig:optics}
\end{figure}

The radiation from the secondary mirror is imaged onto the focal plane via three separate optical
tubes, one for each frequency band; each tube contains an independent system of cold
silicon lenses, a cold Lyot stop, and frequency filters (Fig.~\ref{fig:optics}).  ACT collaborators have developed
high-performance anti-reflection coatings for the silicon lenses \cite{lau06} to reduce internal
reflections to an acceptable level. Three separate bolometer arrays, described
below, lie in the focal plane to detect the radiation. Detecting three frequency bands simultaneously
has the advantages of eliminating different time-dependent systematic effects between the bands,
especially atmosphere, and increasing the efficiency of observations; the price is a
larger receiver. The three detector arrays are arranged so that approximately
75\% of the sky area covered by at least one detector array will be covered at all three frequencies.

The bolometers operate at superconducting temperatures below 1 K; the entire optical assembly,
focal plane, and detector assembly is contained in a large cylindrical dewar approximately
1 meter in diameter and 2 meters in length.  Two stages of pulse-tube coolers maintain the outer optics at 40 K and an inner region at 3 K; helium-4 and helium-3 sorption refrigerators bring the temperature of the detectors down to under 300 mK. The entire refrigeration system contains no consumable liquid cryogens.

As of mid-2006, telescope construction is nearly complete, and the instrument is undergoing
engineering tests. During the second half of 2006, the telescope will be partially disassembled
and shipped to Chile. It will then be reassembled and installed at its observing
site on Cerro Toco, at an elevation of 5100 meters. The telescope will have the capability
of remote operation from a base in San Pedro, Chile.

\section{Detectors}

\begin{figure}[!t]
\includegraphics[width=13.8cm]{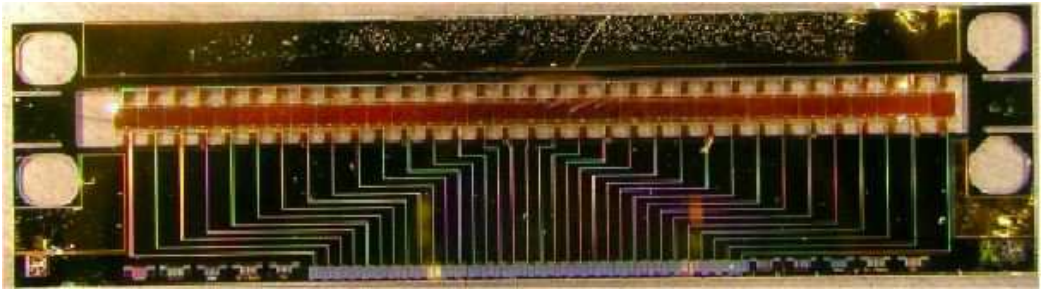}
\caption{The layout of a one-column array of 32 TES bolometers, with leads visible
in the lower half. The row of 32 square bolometers is visible across the upper-center of
the photograph. This array will be folded so that the row of bolometers will sit in the
focal plane, while all other structures seen here will be perpendicular to them and
form part of the mechanical support. Each bolometer is 1 mm square.}
\label{fig:bolometer}
\end{figure}

Large detector arrays are
necessary to reach the sensitivity goals for ACT and similar experiments: photon
shot noise sets a fundamental noise limit for any single detector of microwave background radiation flux.
The heart of the ACT experiment are the bolometer arrays and related electronics,
called the Millimeter Bolometric Array Camera (MBAC). This detector design has been
presented in some detail elsewhere \cite{mar06,nie06} and will only be summarized here.
When the MBAC is complete, it will use three $32\times 32$ arrays of popup bolometers,
developed at NASA Goddard Space Flight Center. The
arrays are divided into columns of 32 bolometers; each column is modular and is
fabricated as a single unit. The bolometers are each squares with side length 1 mm, and
each array is closely packed. Each column of
bolometers and associated leads and support structures are etched onto silicon using lithographic techniques (pictured in Fig.~\ref{fig:bolometer}). The silicon sheets are then folded
in a custom-designed jig so that all of the visible structures above and below the bolometers themselves
are perpendicular to the bolometer row. The bolometers lie in the focal plane; the perpendicular
structures physically support the bolometers and contain electrical leads.

The MBAC uses transition edge-sensing (TES) bolometers \cite{irw96,lee96}. To increase
sensitivity, the bolometers are cooled to their superconducting transition temperature,
and a fixed bias voltage is applied across them. When the amount of radiation falling
on the detector changes, its temperature changes slightly; since it is at its superconducting
transition, a tiny temperature change translates into a large change in resistance. This
results in a changing current through the bolometer, which is detected with
a superconducting quantum interference device (SQUID) via magnetic inductance. 
Multiplexed arrays of these readout SQUIDs have been developed at the National
Institute of Standards and Technology \cite{che99}, reducing the required number of connections from
the cold detectors to the room-temperature data acquisition electronics by a factor of ten.
The detectors must have extensive magnetic shielding since the SQUIDs are highly sensitive to changing magnetic fields.

\section{Observations}

\begin{figure}[!t]
\includegraphics[width=13.8cm]{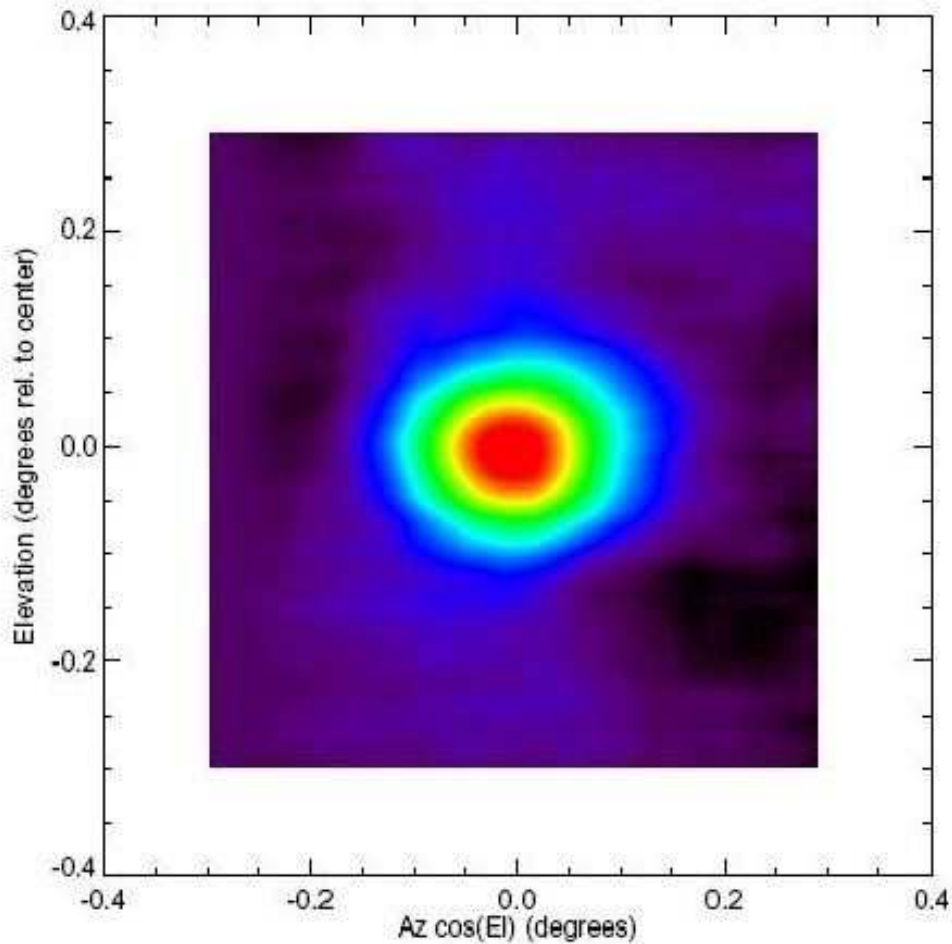}
\caption{This image of Saturn at 145 GHz was made with a multiplexed TES-bolometer
MBAC prototype detector. 
The outputs of 22 detectors have been coadded.
A cryogenic neutral density filter attenuates the input to enable these observations
from Princeton, New Jersey. A Lee spatial filter has been applied.
These data were taken in November, 2005.}
\label{fig:saturn}
\end{figure}

A prototype detector consisting of a single column of 32 multiplexed TES bolometers was
coupled to the sky using a spare mirror for the WMAP satellite and used to observe the
sky from the physics building at Princeton University in late 2005. A resulting image of Saturn
at microwave frequencies is shown in Fig.~\ref{fig:saturn}. This follows the first multiplexed
TES bolometer sky measurements made several years ago \cite{ben02}. 
The first MBAC array is currently being
constructed; a partial array will be used for the first observations with ACT from Chile. 

The Atacama site on Cerro Toco is at the location of the MAT/TOCO experiment \cite{mil02}, and is near
the ALMA, APEX, and CBI experiments. The atmosphere is very dry, minimizing
water vapor absorption; extensive atmospheric data at millimeter wavelengths 
has been taken in preparation for
ALMA.  The site is at a latitude of -23$^\circ$. Azimuthal scans 
at a constant elevation near 45$^\circ$ will map the sky around a strip centered on
a particular declination, with scans taken to the east and west of the South pole
crossing at a nearly perpendicular angle. The initial ACT survey region will be 
a strip roughly two degrees wide centered at declination -55.2$^\circ$, covering the entire
24 hours in right ascension. This particular  declination was chosen to maximize the
strip area with very low galactic dust emission. The strip also contains a number of
known X-ray clusters, including the heavily studied ``hot bullet cluster"
1E0657-56 \cite{mar04,clo04}, the Marano Field which has been imaged deeply
at a range of frequencies \cite{zam99,gru97,elb99,gie03}, and three radio point sources detected by WMAP \cite{ben03b}. The galactic plane
intersects our strip in two locations, and this region of the strip will be unsuitable for cosmological 
observations due to excessive dust emission. The best region of the strip is from RA 22 hr
to 7 hr.

When ACT observations begin from Chile, the telescope will be run continuously,
mostly doing the same two constant-elevation scans, centered on each side of the
South celestial pole.  The bolometer signals will be sampled at around 100 Hz,
resulting of a data rate on the order of one MB per second, or on the order of
100 GB per day.  The time stream data will include the constant sky signal,
time-varying instrument noise, and time-varying but largely coherent atmospheric signal.
With thousands of detectors, ACT will have highly redundant information, 
and the various cosmic and instrumental signals can be reliably separated through their different
temporal and spatial dependences. A given point on the sky will be sampled on a range of time
scales: the pixel transit time during a scan, the array transit time, the scan rate, the pixel sky drift time,
the array sky drift time, the time between east and west scans of the same point, and
the day time scale for observing the entire region. This hierarchy of observation time scales
allows detection and elimination of numerous systematic effects.
Techniques for inverting the time stream into sky maps are
well known, but making maps of reasonable accuracy in close to real time as the data is obtained is
computationally non-trivial, especially if the noise correlations between different detectors
are significant. 

Our baseline plan is to make engineering observations starting at the end of 2006, then 
perform science observations for two to three months in each of  years 2007 and 2008.
The total amount
of data collected during these runs will be in the range of 15 to 20 terabytes. 

\section{Sky and Instrument Simulations}

We have constructed instrument and sky simulations for testing
analysis pipeline code. The instrument simulator includes an accurate model of
the MBAC focal plane, detector thermal and $1/f$ noise and noise correlations
between detectors, detector time constants, model beam shapes, a pointing model,
an atmosphere model, a microwave sky model, and routines to produce a model
data time stream. The atmosphere model is currently a simple frozen turbulence
model \cite{lay00} which is blown across the field of view with a constant velocity; more sophisticated
models are simple to incorporate. 

The sky model includes primary blackbody microwave background fluctuations,
thermal and kinematic SZ effects taken from a large cosmological N-body simulation
with a gas physics prescription \cite{ost05}, relativistic corrections to the thermal SZ effect \cite{noz98},
galactic dust \cite{sch98}, and a random distribution of radio \cite{kno04} and infrared
\cite{bor03}  point sources with varying
spectral indices based on observations. Further map iterations will also include
gravitational lensing by large-scale structure and the Rees-Sciama Effect. Additional
signals which are not yet modeled but can be added are the SZ signal
from individual galaxies and Ostriker-Vishniac fluctuations induced during reionization.
The sky maps are constructed for a strip several degrees wide and at constant
declination, mirroring the ACT sky region. We have produced simulated maps at
the three ACT frequency bands, plus a map at 90 GHz which is useful for studying
relative calibration with WMAP; all of these maps are at resolutions of 0.25 arcminutes. 
 These maps can be used for evaluating the
impact of various experimental noise sources on sky maps reconstructed from the
data, determining how well small map signals can be extracted from noisy data, and 
modelling the ACT galaxy cluster selection function. The simulated sky maps are available 
upon request;
a cluster catalog from the large-scale structure simulations used to make the maps is also available. 

\section{Complementary Observations}

Exploiting fully the cosmological information in high-resolution microwave background observations requires complementary information from other wave bands. In particular, galaxy clusters which
are efficiently imaged via their thermal Sunyaev-Zeldovich signature, independent of their
distance, require optical follow-up imaging to establish redshifts.  This galaxy cluster redshift
catalog, complete down to a given SZ distortion, forms the basis for cosmological tests and probes
of cluster evolution.  For standard cosmological models, ACT will detect over
one thousand galaxy clusters above a mass limit of $2\times 10^{14}\,\,M_\odot$ over a sky
area of 200 square degrees. Thus followup at other wavelengths requires significant telescope
resources and data analysis. 

Several ACT collaborators are part of the Blanco Cosmology Survey (BCS) project, led by
Joe Mohr \cite{bcs}. This effort is imaging 100 square degrees of sky in four optical bands 
(SDSS g, r, i, z) using the Blanco Telescope at CTIO. The first season of data has yielded approximately
30 square degrees of imaged sky, most in four bands. The survey covers two fields,
one centered around RA 23 hr and the other around RA 5.5 hr; the former field will cover
less range in declination and more in RA, compared to the roughly square 5.5 hour field. 
Both extend to a southern declination of around -56.5$^\circ$, so that the ACT strip runs
along their southern edges. Four-band imaging will provide moderately good
photometric redshifts for galaxies, and provide imaging for planning spectroscopic
follow-up observations of galaxy clusters. BCS collaborators are also working on
extracting shear maps from the galaxy images, making use of the Deep Lens Survey
pipeline which was constructed for the same telescope and detector. Two further blocks
of observing time will occur at the end of 2006 and the end of 2007.  The Blanco Cosmology
Survey is a joint project involving three leading groups working on making high-resolution
microwave temperature maps (ACT, the South Pole Telescope, and the Atacama Pathfinder
Experiment). 

ACT also plans to use the new Southern African Large Telescope (SALT) for extensive
optical observations \cite{salt}. 
ACT collaborators at Rutgers and University of KwaZulu-Natal (South Africa) 
have committed 20 nights per year on this 11-meter telescope to support ACT. SALT has
an imaging camera with an 8 arcminute field of view, and a spectrograph with imaging Fabry-Perot
and multi-slit modes of operation. We plan to do multi-band imaging of the ACT region which
is not covered by the Blanco Cosmology Survey, but SALT's primary advantage will be in
obtaining spectroscopy of galaxies in clusters. The multi-object mode of the spectrograph
will be able to obtain 50 to 100 galaxy spectra simultaneously, and the 11-meter aperture
of the telescope makes spectroscopy of faint high-redshift galaxies efficient. Spectroscopy
will provide dynamical galaxy cluster mass estimates and constraints on stellar populations, along
with accurate cluster redshifts. SALT had its
official first light in November 2005, but is still in an engineering observation mode. 

ACT collaborators are involved with the Astronomical Thermal Emission Camera (AzTEC),
a 144-bolometer detector optimized for infrared observations. AzTEC successfully observed 
the sky from the James Clerk Maxwell Telescope in 2005; it is scheduled to be placed on the
Large Millimeter Telescope in Mexico, which can observe the ACT region, in 2007. 
High redshift galaxies with large amounts of
dust emission are common and are a significant foreground in the ACT frequency bands.
Further observations are necessary to gain a better
understanding of this population and its possible impact on cluster Sunyaev-Zeldovich 
measurements. 

Radio point sources, some of which appear in the WMAP data, are another possible
complication for ACT. We are pursuing cluster observations with the Very Large Array
to investigate the extent to which radio sources are correlated with galaxy clusters. We
are investigating possibilities for a radio survey of the ACT region to identify and remove
the brightest radio sources from the ACT maps. 


\section{Institutions and Support}

The Atacama Cosmology Telescope project has drawn together collaborators 
with wide-ranging skills from many US and international institutions. Around 30 faculty-level scientists are actively involved, with about 20 postdocs and graduate students. Princeton University is the lead institution and Lyman Page of Princeton
is the Principal Investigator. The collaboration also includes members from Cardiff University (UK),
Catolica University (Chile),
Columbia University, Haverford College, Lawrence Livermore National Lab, NASA Goddard
Space Flight Center, National Institute for Astrophysics, Optics, and Electronics (INAOE, Mexico),
National Institute of Standards and Technology, Rutgers University, University of British
Columbia (Canada), University of KwaZulu-Natal (South Africa), University of
Massachusetts, University of Pennsylvania, University of Pittsburgh,
University of Toronto (Canada), and York College of the City University of New York. 

ACT is funded primarily by National Science Foundation grant AST-0408698,
and is also supported by NSF grant PHY-0355328. A substantial
amount of support comes from participating institutions in the form
of in-kind contributions, telescope time, and computing resources. 
Additional support for related astronomical observations and educational
programs comes from the NSF's Program in International Research and
Education grant OISE-0530095. AK gratefully acknowledges support by 
NSF grant AST-0546035.
Michael Niemack, Aurelien Fraisse, and 
Lyman Page provided helpful comments about an earlier draft
of this paper.

\end{document}